%% file: main.tex
\documentclass[%
 reprint,
 amsmath,amssymb,
 aps,
 prb,
 floatfix,
]{revtex4-1}

\usepackage{graphicx}
\usepackage{dcolumn}
\usepackage{bm}
\usepackage[english]{babel}

\begin{document}
\title{Commensuration Effects in Layered Nanoparticle Solids}

\author{Luman Qu$^1$, Chase Hansen$^1$, M\'arton V\"or\"os$^2$, and Gergely T. Zimanyi$^1$}
\affiliation{$^1$ Physics Department, University of California, Davis}
\affiliation{$^2$ Materials Science Division, Argonne National Laboratory, Lemont, IL 60439}

\date{\today}

\begin{abstract}
We have developed HiNTS, the {\bf Hi}erarchical {\bf N}anoparticle {\bf T}ransport {\bf S}imulator, and adapted it to study commensuration effects in two classes of Nanoparticle (NP) solids: (1) a bilayer NP solid (BNS) with an energy offset, and (2) a BNS as part of a Field-Effect Transistor (FET). HiNTS integrates the ab initio characterization of single NPs with the phonon-assisted tunneling transition model of the NP-NP transitions into a Kinetic Monte Carlo based simulation of the charge transport in NP solids. First, we studied a BNS with an inter-layer energy offset $\Delta$, possibly caused by a fixed electric field. Our results include the following. (1) In the independent energy-offset model, we observed the emergence of commensuration effects when scanning the electron filling factor $FF$ across integer values. These commensuration effects were profound as they reduced the mobility by several orders of magnitude. We analyzed these commensuration effects in a five dimensional parameter space, as a function of the on-site charging energy $E_C$, energy offset $\Delta$, the disorder $D$, the electron filling factor $FF$, and the temperature $k_{B}T$. We demonstrated the complexity of our model by showing that at integer filling factors $FF$ commensuration effects are present in some regions of the parameter space, while they vanish in other regions, thus defining distinct dynamical phases of the model. We determined the phase boundaries between these dynamical phases. (2) Using these results as a foundation, we shifted our focus to the experimentally much-studied NP-FETs. NP-FETs are also characterized by an inter-layer energy offset $\Delta$, which, in contrast to our first model, is set by the gate voltage $V_G$ and thereby related to the electron filling $FF$. We repeated many of our simulations and again demonstrated the emergence of commensuration effects and distinct dynamical phases in these NP-FETs. Notably, the commensuration effects in the NP-FETs showed many similarities to those in the independent energy-offset BNS.

\begin{description}

\item[PACS numbers] 73.63.Kv

\item[Keywords]nanoparticle, quantum dot, commensuration, FET, transport
\end{description}
\end{abstract}

\pacs{Valid PACS appear here}
\maketitle

\section{\label{sec:introduction}Introduction}

Colloidal semiconductor nanoparticles (NPs) are singularly promising nanoscale building blocks for fabricating mesoscale materials that exhibit emergent collective properties. There is a growing interest to use NPs for numerous optoelectronic applications\cite{talapin_prospects_2010,doi:10.1021/nn506223h}, including third generation solar cells\cite{Nozik02,doi:10.1021/jp806791s} light emitting diodes\cite{shirasaki2013emergence}, and field effect transistors (FET)\cite{Talapin07102005,hetsch_quantum_2013}.

One of the central challenges in all of these applications is to improve the transport in the films, layers, and solids formed from nanoparticles. The value of hopping mobility in today's weakly-coupled insulating NP solids is typically low, $10^{-3}-10^{-2} {\rm cm^2/Vs}$. Various groups attempted to boost the mobility by boosting the inter-NP transition rate with a variety of methods, including: ligand engineering \cite{wang2016colloidal,jang_temperature-dependent_2014,lee_band-like_2011}, band-alignment engineering\cite{chuang2014improved,kroupa2017tuning}, chemical-doping\cite{chen2016metal,choi_bandlike_2012}, photo-doping \cite{talgorn2011unity}, metal-NP substitution\cite{cargnello2015substitutional}, epitaxial attachment of NPs\cite{doi:10.1021/acs.nanolett.6b02382,whitham2016charge}, and atomic layer deposition methods\cite{liu2013pbse}. Encouragingly, these efforts recently translated into progress, as NP films were reported to exhibit band-like, temperature-insensitive mobilities, with values approaching 10 ${\rm cm^2/Vs}$ at room temperatures. 

High conductivities require high mobilities and high carrier densities. However, introducing charge carriers in the nanoparticle solids (NP solids) is challenging. Due to intrinsic difficulties of doping of NPs by impurity atoms\cite{norris2008doped}, so far there have been only a limited number of experimental works achieving successful bulk doping\cite{kang2013influence,sahu2012electronic,mocatta2011heavily}.

Introducing carriers by applying a gate voltage $V_G$ in a field effect transistor (FET) architecture is another promising approach. Several groups reported highly enhanced conductivities in FETs formed from Nanoparticle solids (NP-FETs). \cite{kang2013influence,liu2013pbse}. 

In NP-FETs, a notable issue is the spatial, layer-to-layer distribution of the added carriers. Mean-field analyses of the electron density and the conductivity of FETs\cite{liu2013pbse,hetsch2013quantum}, including Debye-Huckel estimates, as well as detailed experiments on PbSe NP-FETs\cite{mattlawunpublished}, all conclude that electrons, introduced to the layered NP solid by the gate voltage $V_G$, occupy only the first couple layers closest to the gate\cite{reich2014theory}.

This strongly confined spatial distribution of the carriers has the potential to profoundly effect the mobility and thus the conductivity. Most notably, it can lead to commensuration effects via the Coulomb-blockade mechanism. Such commensuration effects have been observed experimentally in CdSe NP solids \cite{houtepen2005orbital}, PbSe NP solids\cite{romero2005coulomb}, and Si NP solids \cite{leobandung1995observation}, among others. 

Besides the obvious scientific interest in understanding the physics of commensuration, it is imperative to get these effects under control for optimizing NP-FETs for technical applications, as Coulomb blockade effects can substantially reduce or even zero out transport.
 
Important early steps in this direction were reported in the recent work of the Shklovskii group. \cite{reich2014theory,reich2015accumulation} They analyzed the non-trivial evolution of the electron distributions in the first and second layers, and the resulting low-temperature conductivity, as the overall electron filling was varied. One of the key outcomes of this work was the theoretical demonstration of strong commensuration effects emerging. They were driven by the complex interplay of the long range Coulomb interaction and the other energy scales of the problem.

This important work was our motivation to explore the physics of commensuration in NP Solids. We focused our analysis on two previously unexplored directions. First, experimental evidence strongly suggests that the screening of Coulomb interactions is strikingly efficient in NP-FETs.\cite{kang_size-_2011} In some cases, the screening by the embedded NPs can be represented by a dielectric constant $\epsilon$ of the order of 10 or higher. Therefore, at least classes of NP-FETs are probably more faithfully modelled by concentrating on the short ranged, "on-site" Coulomb charging energy $E_c$, instead of keeping the entire long range form. This position is supported by the observed temperature dependence of the conductivities: at low temperatures, experiments often report Efros-Shklovskii type variable range hopping, pointing to the importance of keeping the long range part of the Coulomb interaction, whereas above $T\approx 50-80K$, the Efros-Shklovskii temperature dependence typically gives way to a simple activated form, suggesting that the long range portion of the Coulomb interaction ceases to be crucial. Obviously, for solar and optoelectronic applications this second, higher temperature range is of primary interest.

Second, earlier papers did not concentrate on the mobility as a function of the electron filling $FF$, an experimentally relevant parameter, potentially tunable by the gate voltage $V_G$ in NP-FETs. Instead, they studied the $1p^e-1s^e$ energy splitting as the energy scale competing with Coulomb phenomena. In some NP-FETs, such as PbSe NP-FETs, this splitting can be as high as 200 meV, and thus may not be activable at the temperatures of interest. For both of these reasons, we expressly introduced the gate voltage $V_G$ into our model, while dropping the representation of the $1p^e$ energy levels.

In this paper, we adapt our previously developed Hierarchical Nanoparticle Transport Simulator (HiNTS) code to model bilayer NP solids (BNSs). 
HiNTS integrates the ab initio characterization of single NPs with the phonon-assisted tunneling transition model of the NP-NP transitions into a Kinetic Monte Carlo based simulation of the charge transport in NP solids.

Our main results include the following. (1) Starting with the model having an independent inter-layer energy offset $\Delta$, (1.1) we observed the emergence of commensuration effects when the electron filling factors $FF$ in both NP layers reached integer values. These commensuration effects were profound and consequential as they reduced the mobility by orders of magnitude. This reduction is much more substantial than the mobility reductions observed in the long range interaction case. (1.2) We showed the complexity of our model by demonstrating that different classes of commensuration effects emerge in different parameter regions, defining distinct dynamical phases. (1.3) We studied these commensuration effects in a five dimensional parameter space, as a function of the on-site charging energy $E_C$, the energy offset $\Delta$, the disorder $D$, the electron filling factor, $FF$, and the temperature $k_BT$. We explored the dynamical phases in this 5D parameter space that were dominated by the different commensuration effects, and the phase boundaries between them.

(2) Second, we built on our independent energy offset model to describe NP-FETs by recalling that the Poisson equation relates the gate voltage $V_G$ and thus the energy offset $\Delta$ to the electron filling factor $FF$. We modeled NP-FETs by implementing this $\Delta$-$FF$ relation, in effect simulating the NP-FETs as a reduced-dimensional subset of the independent $\Delta$ model. We found that commensuration effects analogous to those previously observed in the independent $\Delta$ model also emerged in NP-FETs. This demonstrates the usefulness and paradigmatic nature of our findings in the higher dimensional parameter space. 
A word on terminology. We distinguish between disorder driven Coulomb blockades and filling-driven Coulomb blockades. In a disorder driven Coulomb blockade, the electron transport is reduced by the energy cost of creating an electron-hole pair in systems with any electron filling; the most prominent case being “zero filling”, which corresponds to neutral systems. Suppression of transport by this disorder driven Coulomb blockade at any filling is well known.

In contrast, the filling-driven Coulomb blockade reduces transport only at commensurate fillings because all NPs the hopping electron intends to hop onto are already occupied by another electron that repels it, making the hops energetically unfavorable. In other words: the disorder driven Coulomb blockade is driven by electron-hole attraction, the filling-driven Coulomb blockade is driven by electron-electron repulsion. Our paper focuses on studying filling-driven Coulomb blockades. As mentioned above and demonstrated below, multi-layer NP solids exhibit a filling-driven Coulomb blockade in some regions of our five-dimensional parameter space, while in other regions the filling-driven Coulomb blockade is conspicuously absent at nominally commensurate fillings. The emerging dynamical phase diagram is therefore far from obvious and is thus worthy of study.

In some detail, in the simplest one-layer model, the suppression of transport by the filling-driven Coulomb blockade is natural. However, in our more complex bilayer NP Solid model, (a) the electrons can redistribute between the layers, thus \textit{de facto} changing the fillings in each layer, and (b) the disorder can help the electrons to overcome the Coulomb barriers, and inter-layer offsets. In this more complex model, it is far from obvious where the blockaded regions will be located in the five-dimensional model-parameter space.

\section{\label{sec:methods} Simulation Methods}

Recently, we have developed HiNTS, the {\bf Hi}erarchical {\bf N}anoparticle {\bf T}ransport {\bf S}imulator, as a multi-level Kinetic Monte Carlo computational platform, to study transport in nanoparticle solids. Previously, we have used HiNTS to study transport in NP-FETs \cite{carbone_monte_2013}, the metal-insulator transition in NP solids \cite{qu2017metal}, and binary NP solids systems \cite{submitted}. 
For the present study, we have extended HiNTS and introduced new features, in order to study the commensuration effect in NP solids, in particular NP-FETs. The presentation and discussion of our results requires a brief description of the hierarchical levels of HiNTS. The details of our methods are provided in the Appendix.

(1) We adapted a k$\cdot$p calculation of the energy levels of PbSe NPs in the diameter range of 5-7 nm. The theoretical results have been validated via comparison to optical experiments.\cite{kang_electronic_1997} Our model also included the electron-electron interaction on the level of on-site/self-charging energy. This self-charging energy can be calculated by a variety of methods, including the semi-empirical pseudopotential configuration interaction method of Zunger and coworkers\cite{PhysRevLett.73.1039,Zunger-charging} and the tight-binding based many body perturbation theory method of Delerue\cite{PhysRevLett.84.2457}. In this paper we report results with the latter approach, because it represents the details of the dielectric screening more realistically.

(2) On the next length scale of the order of 10 nm, we modelled the hopping transitions between neighboring NPs that are separated by twice a ligand length parameter. We incorporated into our model the Miller-Abrahams single phonon-assisted activated hoppings.

(3) On the hierarchically top length scale of hundreds to a few thousand nanometers, we generated an entire solid sample of the NPs, pair-wise coupled via the framework of step (2). We used the event-driven Molecular Dynamics code PackLSD\cite{donev2005neighbor} to obtain close packed (jammed) NP solids. Each sample contained several hundred NPs. We simulated the Bilayer NP Solids (BNS), or NP-FETs, by forming the NP solid in a simulation volume with a thickness of about two NP diameter. We determined whether the NPs belonged to the first or the second layer by tracking the z-coordinates of the NPs. To capture the naturally occurring randomness, the NP diameters were picked from a Gaussian distribution. This disorder in the NP diameters translated into a disorder of the NP energies with a width $D$.

(4) Finally, we simulated transport across the BNS/NP-FET by adapting and using our Extended Kinetic Monte Carlo (KMC) code that incorporated activated transitions\cite{Bortz197510,carbone_monte_2013} between all neighboring NPs. We injected electrons into the NP-FET to reach a predetermined electrons/NP density. Our central quantity of interest was the electron mobility. We always made sure that the voltage was sufficiently small to keep our simulations in the linear I-V regime.

The NP-NP separation (controlled by the ligand lengths), and the overall hopping attempt rate prefactor was selected such that the simulated mobilities were consistent with the experimental values, such as those from the Law group\cite{lee_nonmonotonic_2012}. We systematically explored wide parameter regions, including that of temperature, disorder, electron density, and Coulomb interaction. For each parameter set, we simulated at least 40, typically several hundred samples.

\section{\label{sec:results} Results and Discussion}

\subsection{\label{subsec1:results} Independent inter-layer energy offset $\Delta$}

To begin the exploration of the commensuration effects, we simulated a single-layer NP solid. Fig. 1 shows the mobility as a function of the electron filling factor $FF$, for two different, experimentally typical NP diameters of $d=5.6$ nm, and $d=6.5$ nm. It was assumed that the diameter $d$ of each kind of NP had a Gaussian distribution around these mean values. Redoing the \textit{ab initio} calculations of the NP energy levels, this diameter disorder translated into an energy disorder of width $D$. Visibly, the mobility shows a profound commensuration effect as the electron density per NP (e/NP), or filling factor $FF$, approaches integer values. Demonstrating the commensuration effect at integer fillings $FF$ in the single-layer NP solid establishes the reference frame for the rest of our simulation work.

\begin{figure}[htp]
\includegraphics[width=0.9\columnwidth]{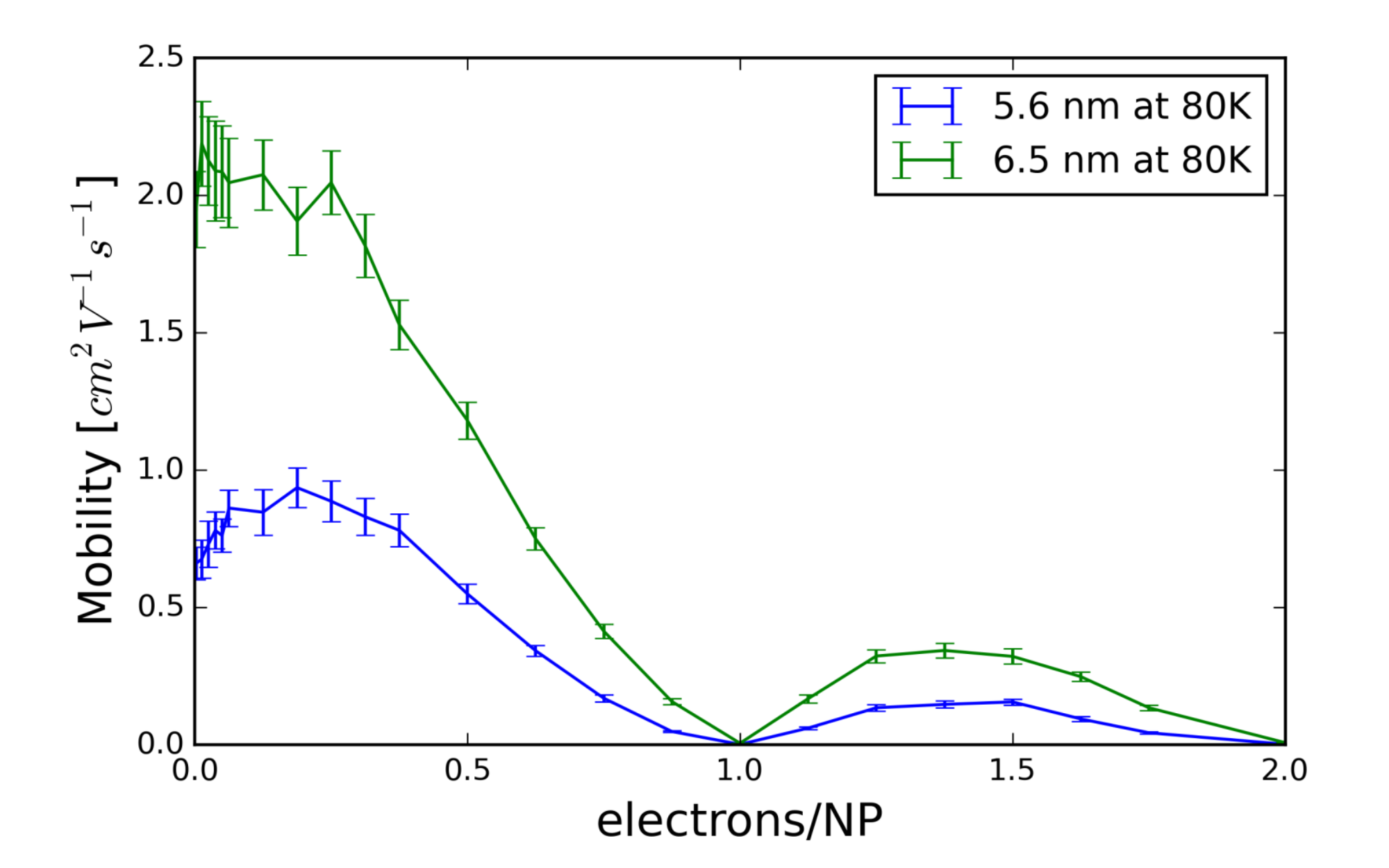}
\caption{Mobility in single layer NP solid, exhibiting a clear commensuration-induced suppression. $E_C=120$ meV, $k_BT=80 K$, $D(d=5.6nm)=55$ meV, $D(d=6.5nm)=45$ meV}
\label{fig:fig1}
\end{figure}

Figs. 2a-b illustrate the underlying physics of this commensuration-induced suppression of the mobility, as the filling $FF$ approaches 1 from below. Fig. 2a shows how the HiNTS code evaluates the energetics of possible transitions for a selected electron (indicated by solid green), when surrounded by NPs that are already occupied by electrons, shown with black. Transition to any site already occupied comes at the additional energy cost of $E_C$, the charging, or on-site Coulomb energy. At low temperatures such energies are not available by thermal assistance, and the selected electron is blocked from executing this transition, as indicated by the red Xs. In the specific case of $FF$ $\rightarrow$ integer, just about all target NPs are already filled with electrons, thus just about all NP-NP transitions are blocked. We refer to this phenomenon interchangeably as the filling-driven Coulomb blockade or commensuration-induced mobility minima. Its primary feature is the exponential suppression of the mobility at integer filling factors $FF$, as shown in Fig. 1, and more compellingly in Fig. 5a.

Fig. 2b illustrates the limits of the commensuration induced by the filling-driven Coulomb-blockade. As the disorder $D$ increases, and becomes comparable to $E_C$, even at commensurate fillings there will be NP-NP transitions where the net energy cost of the transition, of the order of $(E_C-D)$, will become comparable to the thermal energy $k_{B}T$, and thus more and more NP-NP transitions become possible even at $FF=\rm{1}$. This distinction is the basis to define separate dynamic phases of single-layer NP solids: for small disorder $D/E_C<1$, the commensuration induces a Coulomb blockade, separated by a marked transition into a non-blockaded dynamic phase as the control parameter $D/E_C$ exceeds a critical value of the order of 1. We will illuminate this argument with simulations in relation to Fig. 7 below.

\begin{figure}[htp]
\includegraphics[width=1.0\columnwidth]{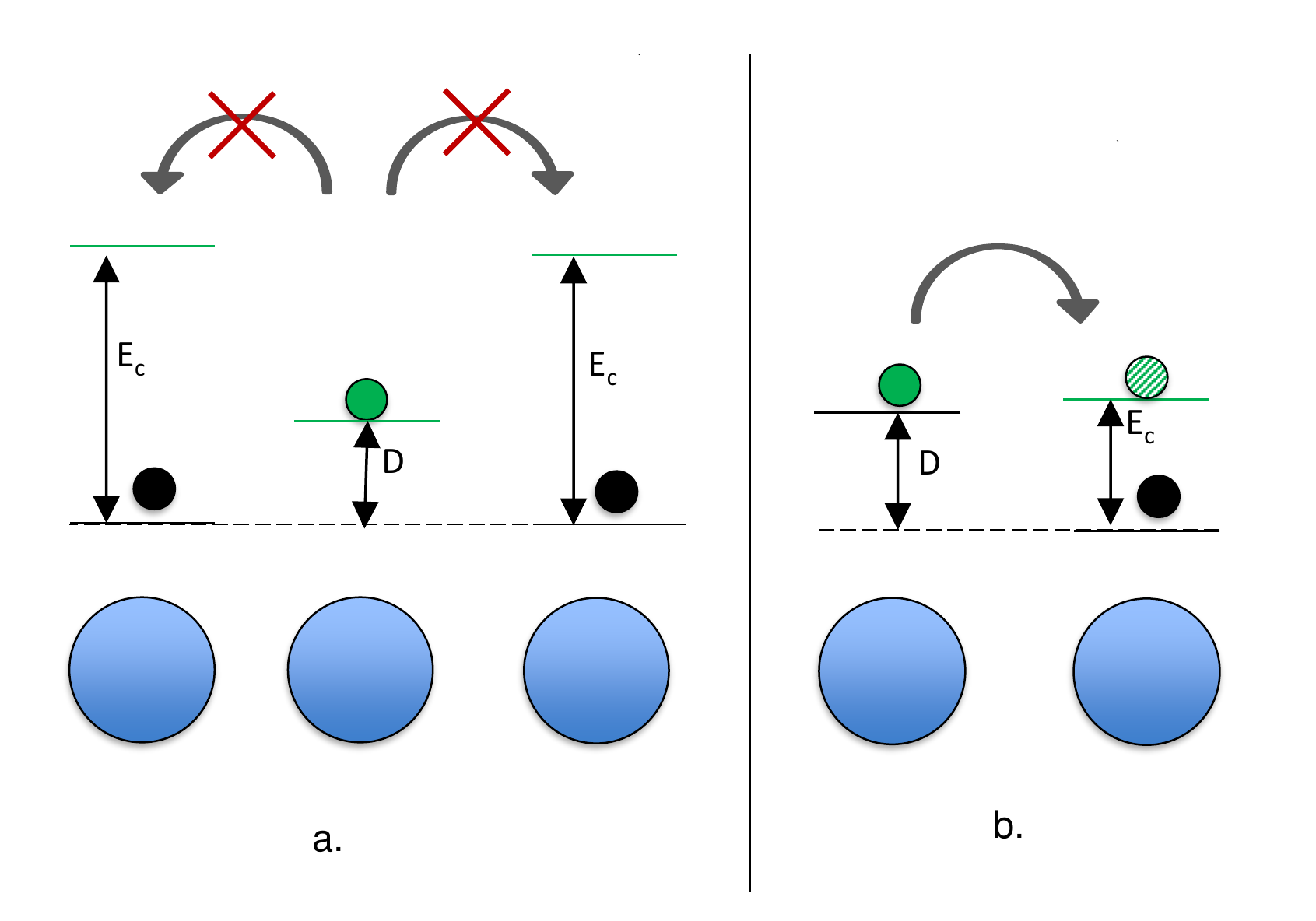}
\caption{The physical mechanism of the Coulomb blockade driving the commensuration-induced suppression of the mobility.}
\label{fig:fig2}
\end{figure}

With this preparation, we now move to the study of Bilayer Nanoparticle Solids (BNS). Fig. 3 illustrates a typical BNS sample. The sample was prepared by PackLSD, as described above in step (3) of HiNTS. The blue/red colors indicate whether a NP belongs to the lower or the upper layer. From here on, the nanoparticles are all selected from a Gaussian distribution of diameters with mean of $d=\rm{6.5nm}$ and a width that translates to an energy disorder of $D$, typically chosen to be $D=\rm{45}meV$. 

\begin{figure}[htp]
\includegraphics[width=1.0\columnwidth]{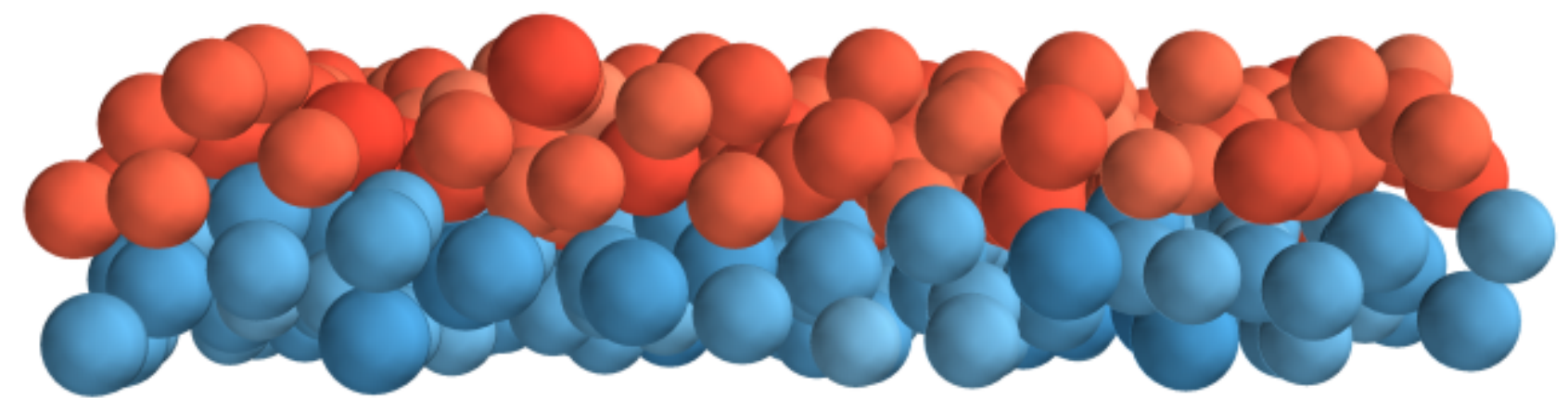}
\caption{Illustration of a simulated bilayer Nanoparticle solid.}
\label{fig:fig3}
\end{figure}

Fig. 4 shows the energy landscape in a BNS. In our model, there is an inter-layer energy offset $\Delta$, which can be caused by various effects, such as a bending of the energy of the conduction band CB or a fixed transverse electric field. This energy offset $\Delta$ is a new competing energy scale in the problem beyond $E_C$ and $D$.

\begin{figure}[htp]
\includegraphics[width=0.6\columnwidth]{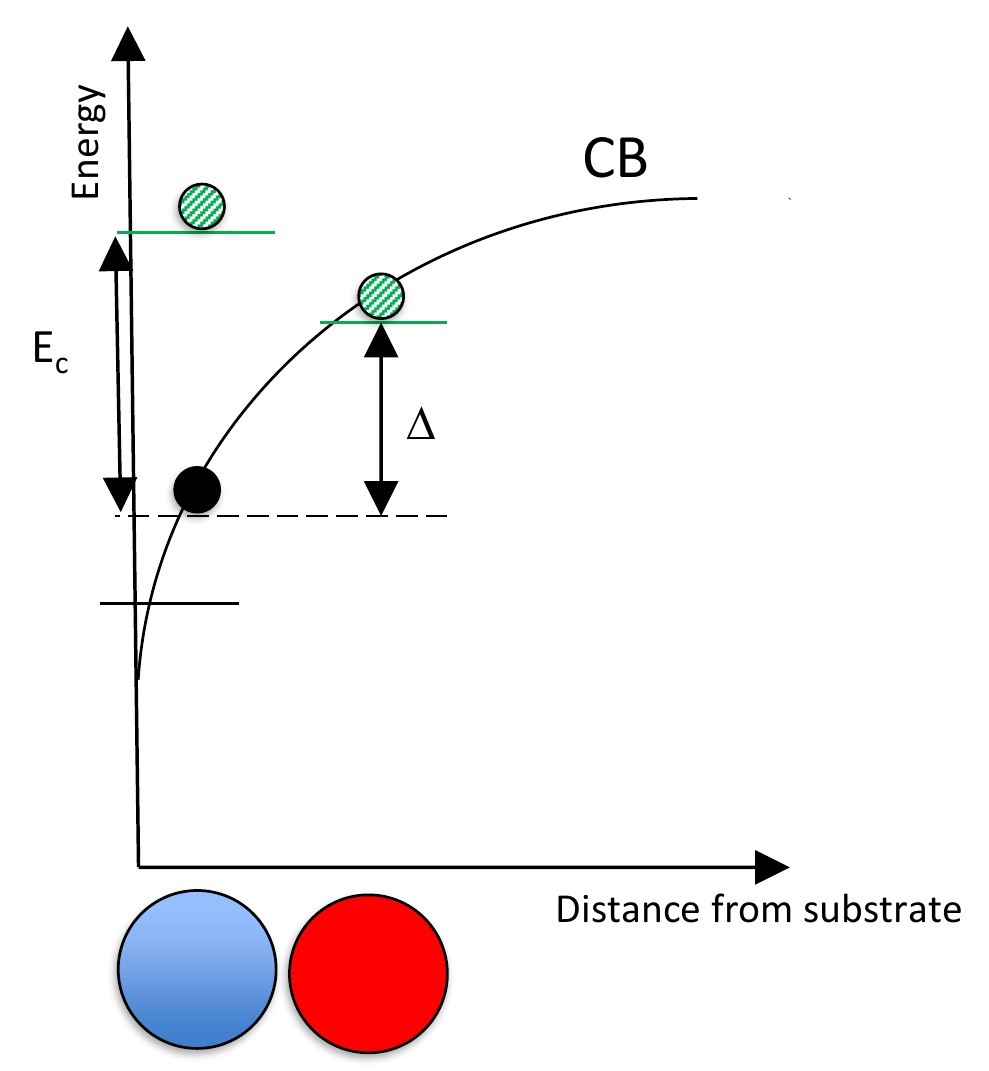}
\caption{Energy landscape of a BNS with an inter-layer energy offset $\Delta$.}
\label{fig:fig4}
\end{figure}

Fig. 5a illustrates the mobility of a BNS as a function of the nominal electron filling factor per layer $FF=e/(NP/layer)$, for different energy offsets $\Delta$. As an example, a BNS in which each layer has 200 NPs, will reach $FF=e/(NP/layer)=1$ when filled by 200 electrons.

\begin{figure}[htp]
\includegraphics[width=1.0\columnwidth]{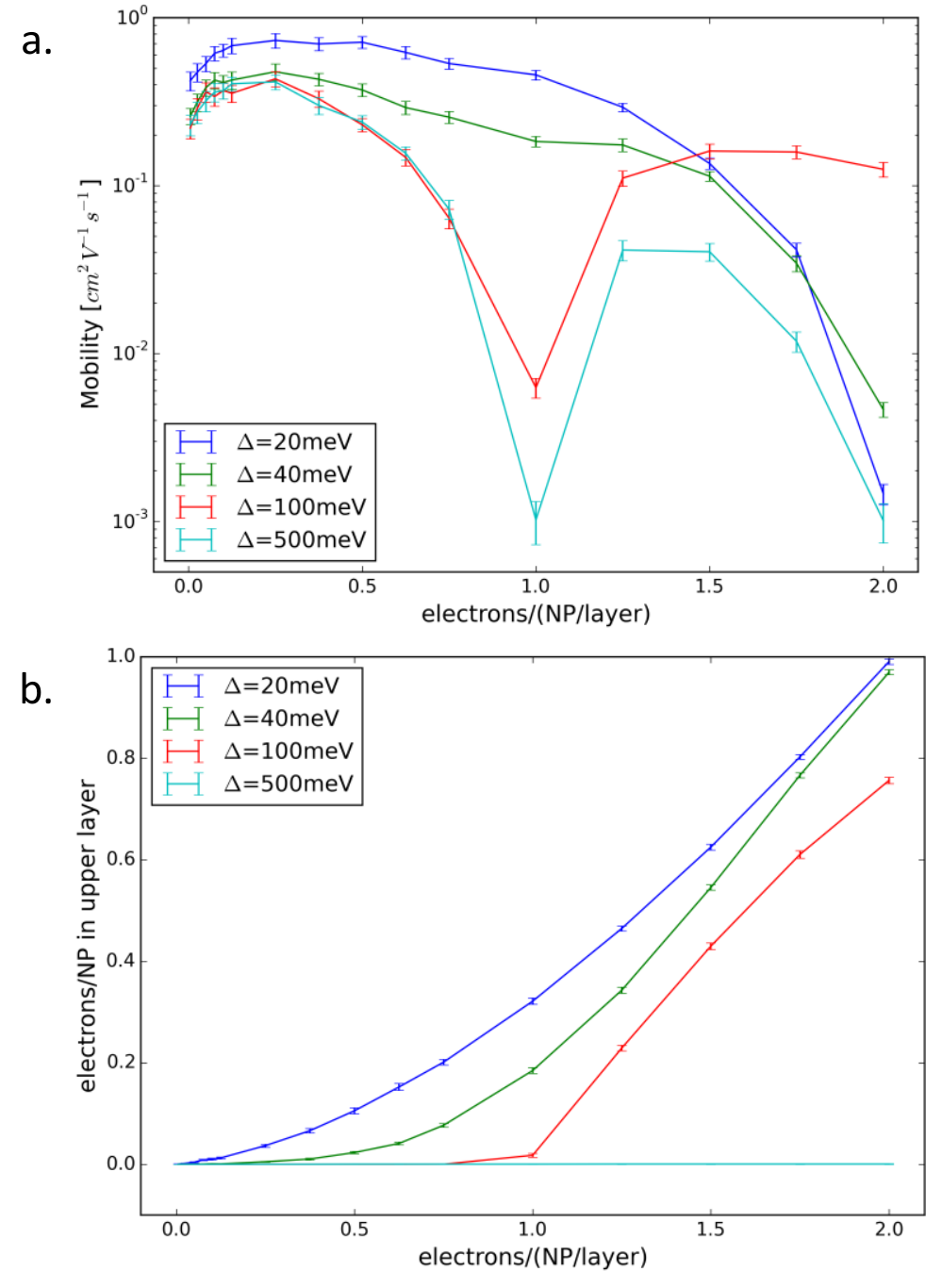}
\caption{Mobility as a function of the filling factor $FF$. $E_C=120$ meV, $k_bT=7$ meV, and $D=45$ meV. Energy offset $\Delta$ is varied from 20 meV to 500 meV.}
\label{fig:fig5}
\end{figure}

Fig. 5a shows that the mobility exhibits profound commensuration-induced minima at $FF=\rm{1}$ for $\Delta=100$ meV and $\Delta=500$ meV, but not at $\Delta=20$ meV and $\Delta=40$ meV; while at $FF=\rm{2}$ surprisingly, for $\Delta=20$ meV, $\Delta=40$ meV, and $\Delta=500$ meV, but not at $\Delta=100$ meV. The log scale shows convincingly that the mobility is exponentially suppressed by 2-3 orders of magnitude relative to the mobilities at non-commensurate $FF$s.

The intriguing complexity of the BNSs is evidenced by the remarkable fact that the commensuration effects emerge at {\it different values of the energy offset} $\Delta$ for $FF=\rm{1}$ and for $FF=\rm{2}$. This is the result of the multi-dimensional competition of the energy scales, as explained next.

To set the stage, Fig. 5b shows $FF_{upper}$, the electron/NP Filling Factor specifically for the upper layer of the BNS, as a function of $FF$, the filling factor of the overall BNS. For $FF\leq 1$, for $\Delta=100$ meV and $\Delta=500$ meV, $FF_{upper}=0$ up to $FF=\rm{1}$, i.e. all electrons remain in the lower layer up to $FF=\rm{1}$.

For $FF\geq1$, $FF_{upper}$ rises for $\Delta=100$ meV, but stays put at $FF_{upper}=0$ for $\Delta=500$ meV. Finally, for $\Delta=20$ meV and $\Delta=40$ meV, $FF_{upper}$ does not show any commensuration effect at $FF=\rm{1}$, but evolves towards the commensurate value $FF_{upper}=1$, as $FF$ approaches 2. These filling commensuration phenomena are summarized in Table I.a.

\begin{table}
	\label{tbl:ff1}
	\begin{center}
		\begin{tabular}{| c | c | c | c | c |} 
			\hline
			$\Delta$ [meV] & 20 & 40 & 100 & 500 \\ 
			\hline
			$FF_{upper}$ & 0.30 & 0.20 & 0.02 & 0 \\ 
			\hline
		    $FF_{lower}$ & 0.70 & 0.80 & 0.98 & 1 \\ 
			\hline
			Commensurate & No & No & Yes & Yes \\
			\hline
		\end{tabular}
	\end{center}
	a. $FF=\rm{1}$
	\begin{center}
		\begin{tabular}{| c | c | c | c | c |} 
			\hline
			$\Delta$ [meV] & 20 & 40 & 100 & 500 \\ 
			\hline
			$FF_{upper}$ & 0.99 & 0.97 & 0.77 & 0 \\ 
			\hline
		    $FF_{lower}$ & 1.01 & 1.03 & 1.23 & 2 \\ 
			\hline
			Commensurate & Yes & Yes & No & Yes \\
			\hline
		\end{tabular}
	\end{center}
	b. $FF=\rm{2}$
	\caption{Summary of commensuration effects at $FF=\rm{1}$ and $FF=\rm{2}$.}
\end{table}
Fig. 6 explains the observations above. Figs. 6a-b are relevant for $FF=\rm{1}$, whereas Figs. 6c-d are relevant for $FF=\rm{2}$. Fig. 6a shows that when the energy offset $\Delta$ is much larger than the disorder $D=45$ meV, for example $\Delta=100$ meV or $\Delta=500$ meV, then, as the electrons are filled into the BNS, they all remain in the lower NP layer. For $FF\leq1$, the electrons do not doubly occupy the NPs, so the charging energy $E_C$ does not enter into the competition of energy scales yet. This explains why $FF_{upper}$ remains zero for $FF\leq1$ for the higher energy offsets of $\Delta=500$ meV and $\Delta=100$ meV. 

Since the competition of $\Delta$ and $D$ confines all electrons into the lower layer, the charging energy $E_C$ induces pronounced commensuration-induced mobility minima at $FF=\rm{1}$ for these high energy offsets, as shown by the blocked NP-NP transition, shown with a red X. Zooming in on the mobility values at the commensuration-induced minima at $FF=\rm{1}$, Fig. 5a shows that the mobility minimum is lower for $\Delta=500$ meV than for $\Delta=100$ meV. This is because $\Delta=500$ meV confines the electrons to the lower layer more effectively, as documented by Table I.a as well.
\begin{figure}[htp]
\includegraphics[width=0.7\columnwidth]{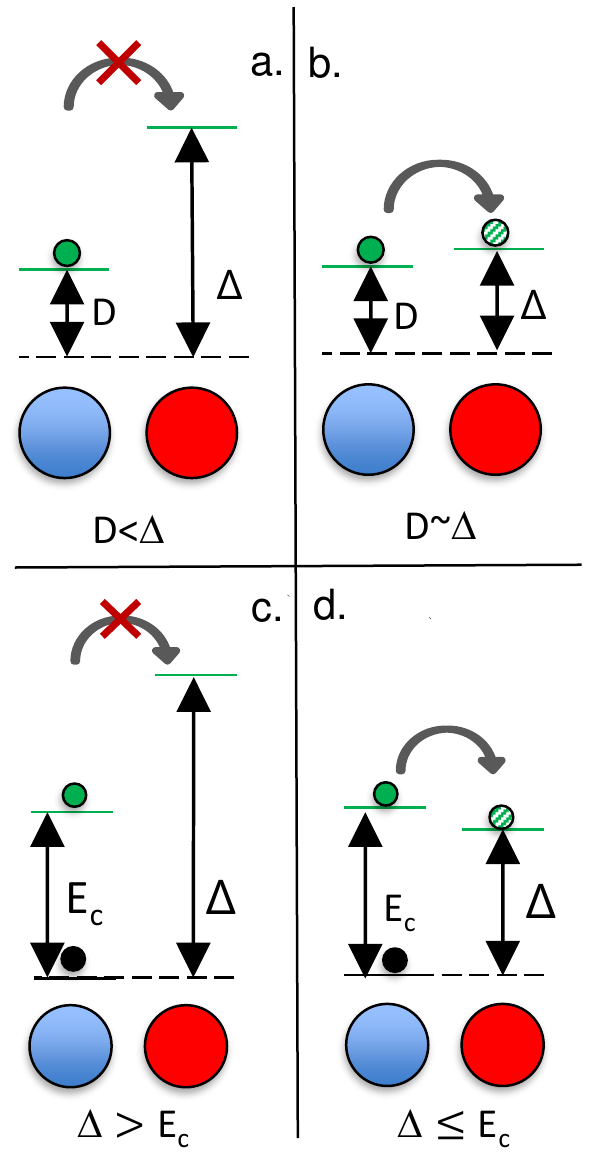}
\caption{Energy diagrams to contextualize the various parameter regimes. (a.) illustrates $FF=\rm{1}$, $\Delta=500$ meV and $\Delta=100$ meV. (b.) illustrates $FF=\rm{1}$, $\Delta=40$ meV and $\Delta=20$ meV. (c.) illustrates $FF=\rm{2}$, $\Delta=500$ meV. (d.) illustrates $FF=\rm{2}$, $\Delta=100$ meV, $\Delta=40$ meV and $\Delta=20$ meV. In all cases $E_C=120$ meV, $k_BT=7$ meV and $D=45$ meV.}
\label{fig:fig6}
\end{figure}
Fig. 6b shows the complementary case of lower, 20 meV and 40 meV values of the energy offset $\Delta$. For these lower offsets, the 45 meV disorder $D$ is capable of overcoming the energy offset $\Delta$ and promoting the electrons from the solid green state on the blue, lower layer NPs to the shaded green state on the red, upper layer NPs, as shown by the allowed NP-NP transition. The possibility of freely transitioning between the lower and upper NP layers increases $FF_{upper}$ to non-zero values, thus making the electron density non-integer in both NP layers. Since only integer Filling Factors activate the Coulomb blockade, these non-integer filling factors wash out the commensuration-driven mobility minima. This explains the disappearance of the commensuration effect in the blue and green curves of the mobility at $FF=\rm{1}$.

Figs. 6c-d are helpful to analyze how the physics of commensuration changes for $FF=\rm{2}$. For these higher fillings, (blue) NPs in the lower layer are often doubly occupied, as shown. Typical values of the charging energy $E_C$ for isolated NPs are about 120 meV, considerably greater than the disorder. Therefore, the charging energy $E_C\approx 120$ meV replaces the disorder $D\approx 45$ meV, as the primary energy scale competitor to the offset $\Delta$.

Fig. 6c is most relevant for the large energy offset of $\Delta=500$ meV. Here, even an $E_C=120$ meV is not capable of promoting electrons into the upper layer. Therefore, $FF_{upper}$ remains zero even as $FF$ grows from 1 to 2, as confirmed by Fig. 5b. By this mechanism, at $FF=\rm{2}$, the filling factors for both layers reach integer values, $FF=\rm{2}$, and $FF_{upper}=0$, thus the Coulomb blockade once again drives a commensuration-induced effect: an exponentially suppressed mobility minimum.

Fig. 6d shows that the physics changes as the energy offset is reduced to $\Delta=100$ meV. At this value, $\Delta$ is reduced to a level comparable to the charging energy $E_C$, thus freeing up the electrons to transition between layers. Fig. 6d shows that the energy of a (green) electron, residing on a (blue) NP in the lower layer, is lifted by the Coulomb repulsion from a (black) electron on the same NP, making the green electron capable of reaching the shaded green electron state on a (red) NP in the upper layer. Notably, since the energy offset and the charging energy are comparable, the spatial distribution of the electrons spreads out over the two layers. This is captured by the $FF_{upper}$ assuming a non-integer, non-commensurate intermediate value in Table I. This explains why the mobility does not exhibit a commensuration-induced minimum.

Finally, for even lower energy offsets $\Delta=20$ meV and $40$ meV, $\Delta$ is markedly smaller than $E_C$. This not only makes it possible for the electrons to leak into the upper layer, much rather it forces the electrons to do so. This is the driver of $FF_{upper}$ actually reaching 1 as $FF$ approaches 2. Since the filling factors of each layer reach integer values at $FF=\rm{2}$, the Coulomb blockades once again drive commensuration-induced mobility minima, as shown in Fig. 5a.

We note, that the commensuration-induced physics is markedly different for the different cases. For $FF=\rm{1}$, the upper layer does not play any role. For $FF=\rm{2}$ and $\Delta=500$ meV, the NPs in the lower layer are doubly occupied, and the upper layer plays no role. Finally, for $FF=\rm{2}$ and $\Delta=20$ meV and $40$ meV, the upper and lower layer play a largely symmetric role. These regimes are dominated by different physics, and therefore can be identified as different dynamical phases of the BNS. The latter two, for example, are separated by a phase boundary around $\Delta\approx E_C$, where the competing energy scales are comparable. This washes out the commensuration effects, and serves as an effective phase boundary between the dynamical phases, as long as both remain large compared to $D$. The commensuration phenomena for the filling factor $FF=\rm{1}$ are summarized in Table I.a.

The commensuration phenomena for the filling factor $FF=\rm{2}$ are summarized in Table I.b. The complexity of the model is on full display in that the commensuration effects at $FF=\rm{1}$, as shown in Table I.a, are reversed relative to $FF=\rm{2}$ for 3 of the 4 values of $\Delta$, as shown in Table I.b. The primary driver of these reversals is that the energy scale that is the primary competitor of $\Delta$ switched from the disorder $D$ at $FF=\rm{1}$, to the charging energy $E_C$ at $FF=\rm{2}$.

\begin{figure}[htp]
\includegraphics[width=1.0\columnwidth]{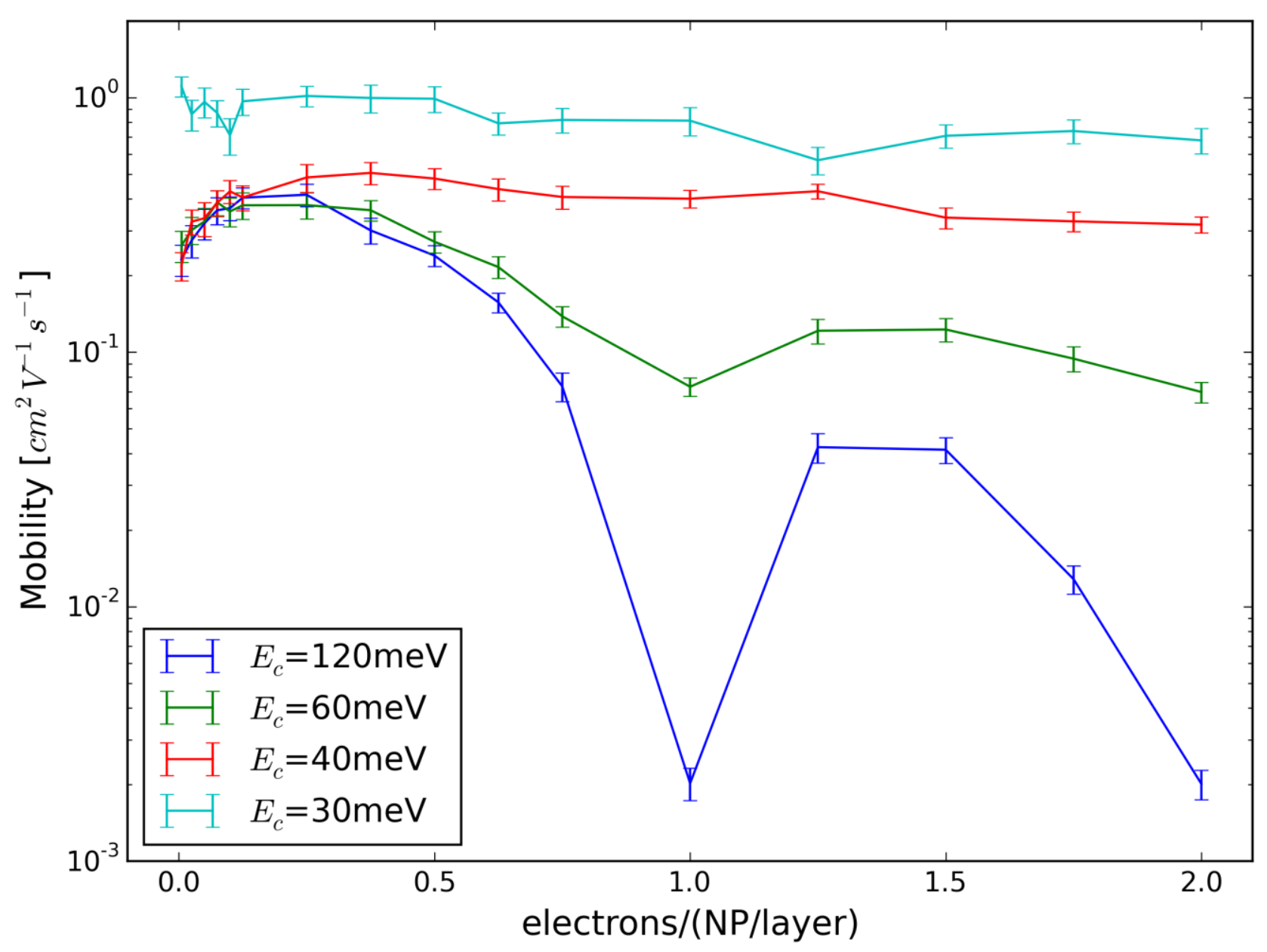}
\caption{Fixed $\Delta$, different $E_C$: $\Delta=500$ meV, $k_BT=7$ meV, and $D=45$ meV.}
\label{fig:fig7}
\end{figure}

Up to now the model was analyzed by scanning the energy offset $\Delta$ and the filling factor $FF$, while keeping the charging energy $E_C$ constant. An informative complementary parameter scan is shown in Fig. 7, where the charging energy $E_C$ is scanned, while keeping the energy offset $\Delta$ constant. In all curves shown, $\Delta=500$ meV. Therefore, all electrons are confined into the lower NP layer, and the physics is determined by the competition of the charging energy $E_C$ and the disorder $D$. The blue curve shows the mobility at $E_C=120$ meV, the same as the lowest curve in Fig. 5a. This parameter set was selected as it shows commensuration-induced mobility minima {\it both} at $FF=\rm{1}$, and at $FF=\rm{2}$. The relevant energy diagram is illustrated in Fig. 2a, showing that the competing energy scale of the disorder $D$ can not help the electrons to overcome the Coulomb blockade either at $FF=\rm{1}$ or at $FF=\rm{2}$. 

As the charging energy is reduced to $E_C=60$ meV, the depth of the mobility minima at $FF=\rm{1}$ and at $FF=\rm{2}$ are greatly reduced. Fig. 2b explains this as follows. At $E_C=60$ meV, the disorder $D$, more precisely, the disorder $D$, augmented by the thermal energy to $D+k_BT$, becomes comparable to $E_C$, and thus capable of boosting the electrons to partially overcome the Coulomb blockade within the first layer. Upon further reduction to $E_C=40$ meV and $30$ meV, the mobility minima are completely smoothed out, as the disorder becomes the dominant energy scale, and the charging energy is unable to hinder transport anymore. 

The above specific scans of the multidimensional parameter space demonstrate that the competition of the main physical processes gives rise to distinct dynamical phases of the model. Next, we create a comprehensive phase diagram of these dynamical phases in the $D$-$\Delta$-$E_C$ space by performing a systematic 2 dimensional raster scan of the $D/E_C$ vs. $\Delta/E_C$ space. At each point of this raster scan we performed a scan with the filling factor $FF$ and determined whether the BNS exhibited a well-defined mobility minimum at the two potential locations of commensuration effects: at $FF={\rm 1}$, or at $FF={\rm 2}$, or both. We adopt a "Dynamic Commensuration Matrix" DCM order parameter to characterize the dynamical phases through their filling factors as follows:

\begin{table}
	\label{tbl:ff2}
	\begin{center}
		\begin{tabular}{| c | c | c |} 
			\hline
			{\bf DCM} & $FF(lower)$ & $FF(upper)$ \\ 
			\hline
			$FF=$ 1 & &\\ 
			\hline
		    $FF=$ 2 & & \\ 
			\hline
		\end{tabular}
	\end{center}
	\caption{Dynamic Commensuration Matrix {\bf DCM} order parameter of the dynamical phases of the BNS model.}
\end{table}
This {\bf DCM} order parameter cross-references the nominal filling factor $FF$ with the actual filling factor $FF(upper)$ of the upper layer, and $FF(lower)$, that of the lower layer at the two potential locations of commensuration effects: at $FF={\rm 1}$, and $FF={\rm 2}$. The top row of the {\bf DCM} order parameter represents $FF(lower)$ and $FF(upper)$ at $FF={\rm 1}$ nominal filling, the bottom row the same fillings at $FF={\rm 2}$.

When the top row of the {\bf DCM} contains the integers (1,0), then the nominal commensuration at $FF={\rm 1}$ indeed induces commensuration in the top and bottom layers, and thus the BNS exhibits commensuration-induced mobility minimum. In contrast, when the top row of the {\bf DCM} contains non-integers, shown in the diagram as (1-n, n), then the competing physical processes smooth out the nominal commensuration, and the BNS does not exhibit mobility minima.

Analogously, when the bottom row of the {\bf DCM} contains the integers (1,1) or (2,0), then the nominal commensuration at $FF={\rm 2}$ indeed induces commensuration in the top and bottom layers, and thus the BNS exhibits a commensuration-induced mobility minimum. In contrast, when the bottom row of the {\bf DCM} contains non-integers, shown in the diagram as (2-n, n), then the competing physical processes smooth out the nominal commensuration, and the BNS does not exhibit mobility minima.

\begin{figure}[htp]
\includegraphics[width=0.9\columnwidth]{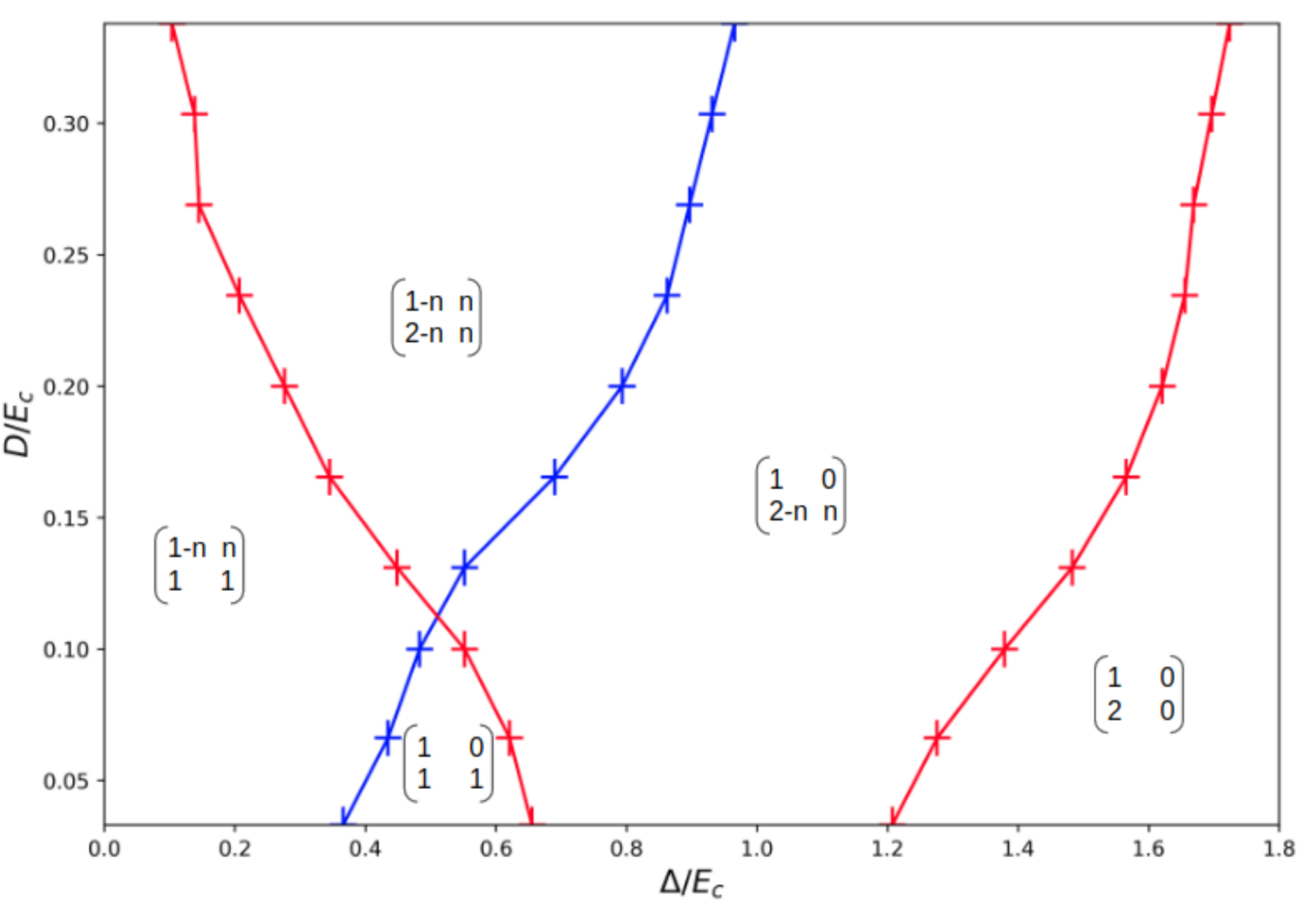}
\caption{Dynamic phase diagram, capturing the dynamics of the BNS at the two filling factors $FF={\rm 1}$ and $FF={\rm 2}$ in terms of the Dynamic Commensuration Matrix {\bf DCM}.}
\centering
\end{figure}

The rich information coded in the {\bf DCM} Dynamical Commensuration Matrix order parameter can be condensed into a simpler Dynamical Commensuration Vector {\bf DCV}. The upper element of the {\bf DCV} only indicates whether at $FF={\rm 1}$ the dynamical phase exhibits a mobility Minimum At Commensuration: "M"; or not: "No-M". The lower element of the {\bf DCV}, only indicates whether at $FF={\rm 2}$ the dynamical phase exhibits a mobility Minimum At Commensuration: "M"; or not: "No-M". With this convention, the five phases of Fig. 8 are the following:

(a) The left-most phase with the lowest $\Delta/E_C$, having a {\bf DCM} = (1-n, n)/(1, 1), where the lower row of the {\bf DCM} matrix is shown after the "/", is described by a {\bf DCV} = (NoM/M), the lower {\bf DCV} vector element also shown after a "/" for consistency. 

(b) The phase with higher $\Delta/E_C$ and higher $D/E_C$, having a {\bf DCM} = (1-n, n)/(2-n, n), is described by a {\bf DCV} = (NoM/NoM).

(c) The phase with similar $\Delta/E_C$ but lower $D/E_C$ (the small upward-pointing triangle based on the $\Delta/E_C = 0.4 - 0.6$ interval), having a {\bf DCM} = (1, 0)/(1, 1), is described by a {\bf DCV} = (M/M).

(d) The phase with yet higher $\Delta/E_C$, having a {\bf DCM} = (1, 0)/(2-n, n), is described by a {\bf DCV} = (M/NoM).

(e) The phase with the highest $\Delta/E_C$, having a {\bf DCM} = (1, 0)/(2, 0), is described by a {\bf DCV} = (M/M).

\begin{table}
	\label{tbl:ff3}
	\begin{center}
		\begin{tabular}{| c | c |} 
			\hline
			{\bf DCM} & {\bf DCV} \\
			\hline
			$\left(
                \begin{array}{cc}
                1-n&n\\
                1&1\\
                \end{array}
            \right)$ & $\left(
                \begin{array}{c}
                NoM\\
                M\\
                \end{array}
            \right)$  \cr
			\hline
		    $\left(
                \begin{array}{cc}
                1-n&n\\
                2-n&n\\
                \end{array}
            \right)$ & $\left(
                \begin{array}{c}
                NoM\\
                M\\
                \end{array}
            \right)$ \cr
            \hline
            $\left(
                \begin{array}{cc}
                1&0\\
                1&1\\
                \end{array}
            \right)$ & $\left(
                \begin{array}{c}
                M\\
                M\\
                \end{array}
            \right)$  \cr
            \hline
            $\left(
                \begin{array}{cc}
                1&0\\
                2-n&n\\
                \end{array}
            \right)$ & $\left(
                \begin{array}{c}
                M\\
                NoM\\
                \end{array}
            \right)$ \cr
            \hline
            $\left(
                \begin{array}{cc}
                1&0\\
                2&0\\
                \end{array}
            \right)$ & $\left(
                \begin{array}{c}
                M\\
                M\\
                \end{array}
            \right)$  \cr
			\hline
		\end{tabular}
	\end{center}
	\caption{The Mapping of the {\bf DCM} Dynamic Commensuration Matrix onto the {\bf DCV} Dynamic Commensuration Vector.}
\end{table}

It is instructive to review the phase diagram from the complementary vantage point of the phase boundaries. 

(1) The blue boundary separates dynamic phases of the BNS that differ by the commensurative behavior at $FF={\rm 1}$, but not at $FF={\rm 2}$. Indeed, for low energy offset $\Delta/E_C$, the electron fillings in the (lower, upper) layers are (1-n, n), thus the BNS is in a non-commensurate dynamic phase that does not exhibit mobility minima at $FF={\rm 1}$. For high $\Delta/E_C$, the electron fillings in the (lower, upper) layers are (1, 0), thus the BNS is in a commensurate dynamic phase that does exhibit mobility minima at $FF={\rm 1}$. Thus, crossing the blue phase boundary by increasing $\Delta/E_C$ is a dynamic phase transition from a phase in which the BNS does not exhibit mobility minima at $FF={\rm 1}$ to a phase in which it does. For completeness, across the blue phase boundary, the commensurative behavior of the BNS does not change at $FF={\rm 2}$.

(2) The red boundary separates dynamic phases of the BNS that differ by the commensurative behavior at $FF={\rm 2}$, but not at $FF={\rm 1}$. Indeed, for low energy offset $\Delta/E_C$, the electron fillings in the (lower, upper) layers are (1, 1), thus the BNS is in a commensurate dynamic phase that does exhibit mobility minima at $FF={\rm 2}$. For medium $\Delta/E_C$, the electron fillings in the (lower, upper) layers are (2-n, n), thus the BNS is in a dynamic phase that does not exhibit mobility minima at $FF={\rm 2}$. Finally, for high $\Delta/E_C$, the electron fillings in the (lower, upper) layers are (2, 0), thus the BNS is again in a dynamic phase that does exhibit mobility minima at $FF={\rm 2}$. Thus, crossing the red phase boundary from low to medium $\Delta/E_C$ is a dynamic phase transition from a phase in which the BNS does exhibit mobility minima at $FF={\rm 2}$ to a phase in which it does not. Further, crossing the red phase boundary from medium to high $\Delta/E_C$ is a dynamic phase transition from a phase in which the BNS does not exhibit mobility minima at $FF={\rm 2}$ to a phase in which it does again. This is an intriguing case of a reentrant phase diagram. Again for completeness, across the red phase boundaries, the commensurative behavior of the BNS does not change at $FF={\rm 1}$.

(3) Broadly speaking, for a given disorder, increasing the energy offset $\Delta/E_C$ at $FF=\rm{1}$ forces more and more electrons into the lower layer, eventually cutting off their ability to escape the Coulomb blockade. This tendency led to the formation of mobility minima at commensuration (M), as it forced all electrons into the lower layer. For $FF=\rm{2}$, the filling of the lower and upper layers evolved from the evenly distributed commensurate $(FF(lower),FF(upper))$ = (1,1) to the moderately uneven and non-commensurate $(FF(lower),FF(upper))$ = (2-n,n), eventually to the fully uneven, commensurate $(FF(lower),FF(upper))$ = (2,0), as $\frac{\Delta}{E_C}$ was increased. 

(4) While we observed that increasing $\Delta/E_C$ induced complex phase transition sequences, from more commensurate to less commensurate, followed by again to more commensurate, the trends with increasing disorder $D/E_C$ were straightforward: more disorder moved the BNS from more commensurate towards less commensurate. Every time when either a blue or red phase boundary was crossed with increasing $D/E_C$ (vertically), commensuration was lost either at $FF=1$, or $FF=2$. The physics behind this is natural: increasing disorder can smooth out the energy differences driven by $\Delta$ or $E_C$, thus smoothing out the mobility minima as well.  

\subsection{\label{subsec2:results} NP-FETs with filling factor $FF$ controlled by the inter-layer energy offset $\Delta_{FET}$}

The independent energy offset model is a well defined statistical physical model, and thus worthy of study. Our extensive exploration created a comprehensive description of the phase diagram. Next, we turn our attention to the specific case of NP-based FETs, which have great potential for applications. We start by recalling that the energy offset $\Delta_{FET}$ is not a free parameter in NP-FETs, since it is induced by the transverse gate voltage, which also impacts the electron filling factor, $FF$ as the two are related via the Poisson equation.

We use Eq. (6) of Shklovskii's 2014 paper\cite{reich2014theory} to represent this relationship. Broadly speaking, in the interval of interest, we take $\Delta_{FET}$ to be proportional to $FF$. Since additionally there are several material parameters in this equation that can vary from FET to FET, we carried out a set of simulations with varying proportionality constants. 

Fig. 9 illustrates our results. We selected a set of proportionality constants between $\Delta$ and $FF$ such that $\Delta_{FET}$ at $FF=\rm{1}$ assumed those values which we used for our fixed-$\Delta_{BNS}$ simulations in Fig. 5a.  This choice created the closest analogy and thus comparability between the two sets of runs. 

While the results in Fig. 9 and in Fig. 5a are not exactly the same, nevertheless they demonstrate the same paradigm. While broadly speaking, the Coulomb blockade tends to suppress the mobility at commensurate electron fillings, whether this suppression actually manifests itself depends in a non-trivial and intricate manner on the various parameters of the model. For some parameters, the NP-FET shows a suppression only at $FF=\rm{1}$, for others only at $FF=\rm{2}$, for some at both fillings, and for some at none at all.

To establish a relationship between the phase diagram of the independent $\Delta_{FET}$ BNS and the NP-FETs, we note that our BNS Dynamical Commensuration Vector {\bf DCV} was defined by the presence or absence of mobility minima at the two filling factors $FF=\rm{1}$ and $FF=\rm{2}$ at the \textit{same} $\Delta_{BNS}$. In contrast, for NP-FETs, the NP-FET Dynamical Commensuration Vector {\bf DCV} is defined by the presence or absence of mobility minima at the two filling factors $FF=\rm{1}$ and $FF=\rm{2}$ using their corresponding, \textit{different} $\Delta_{FET}$s. Thus, the two elements of the {\bf DCV} have to be determined from information in the BNS phase diagram at the two different $\Delta$s that correspond to $FF=\rm{1}$ and $FF=\rm{2}$.

\begin{figure}[htp]
\hspace{-.75cm}
\includegraphics[width=1.0\columnwidth]{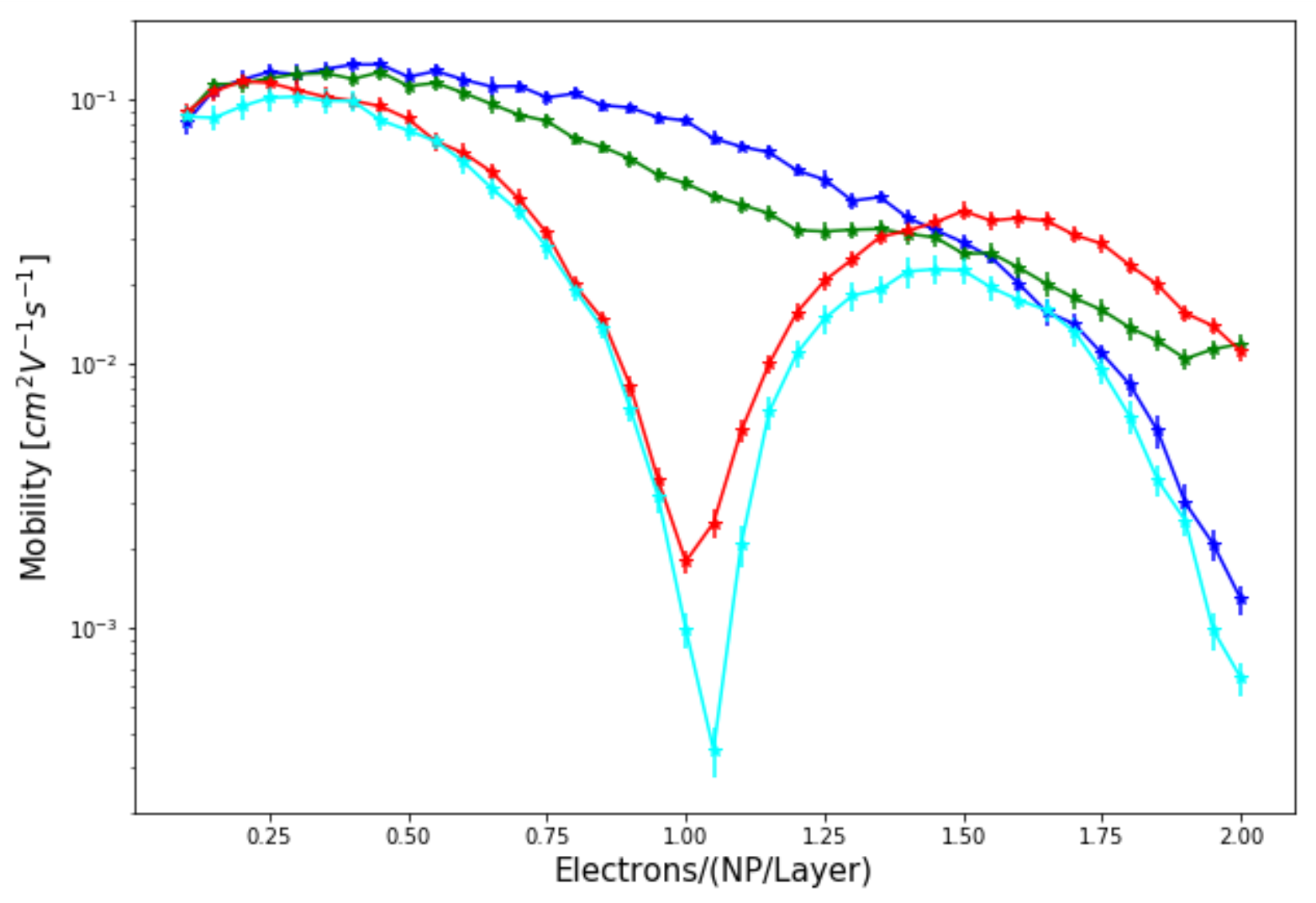}
\caption{Mobility as a function of the electron filling factor $FF$. The 4 curves were generated by using 4 proportionality constants between the inter-layer energy offset  $\Delta_{FET}$ and the filling factor $FF$ such that $\Delta_{FET}$ was equal to the four $\Delta_{BNS}$ values used in Fig. 5a. at $FF=\rm{1}$}


\end{figure}

\section{\label{sec:conclusions} Conclusions}

In this paper, we adapted our previously developed Hierarchical Nanoparticle Transport Simulator (HiNTS) to simulate two models of interest: (1) Bilayer Nanoparticle Solids (BNSs) with an independently variable inter-layer energy offset $\Delta_{BNS}$; and (2) Bilayer NP-FETs, where the inter-layer energy offset $\Delta_{FET}$ controlled the filling factor $FF$. HiNTS combines ab initio single NP modeling and NP-NP transition modeling into a Kinetic-Monte-Carlo-based simulation of the transport in BNSs and NP-FETs. Our main results included the following. 

(1) We observed the emergence of commensuration effects when the electron filling factors in both NP layers reached integer values. These commensuration effects were profound as the on-site Coulomb blockade reduced the mobility exponentially close to zero, often by 2-3 orders of magnitude. These observed reductions are to be contrasted with the limited mobility reductions observed in models with long range interactions. 

(2) We noted that different classes of commensuration effects emerged for different parameter regions. The complexity of the physics was well-demonstrated by the fact that the commensuration effects were markedly different at filling factors $FF=\rm{1}$ and $FF=\rm{2}$: in some regions the mobility showed a minimum at $FF=\rm{1}$ but not at $FF=\rm{2}$, in some cases at $FF=\rm{2}$ but not at $FF=\rm{1}$, in some cases at both, and in some cases at neither. 

(3) We systematically swept a two dimensional subspace ($\frac{\Delta}{E_C}$, $\frac{D}{E_C}$) of the parameter space to construct model's dynamical phase diagram. We introduced two order parameters, the Dynamical Commensuration Matrix {\bf DCM}, and the Dynamical Commensuration Vector {\bf DCV} to capture the presence or absence of mobility minima at $FF=\rm{1}$ and $FF=\rm{2}$ as the electron filling $FF$ was swept.  

We identified five separate dynamical phases of the model that demonstrate the richness of the emergent physics, driven by the competition of the several energy scales of the model. We developed an explanation for the presence or absence of mobility minima in each of these dynamic phases. We developed further insights by discussing the critical behavior as the various phase boundaries were crossed. Finally, we demonstrated the paradigmatic nature of our dynamical phase analysis by reporting that the patterns of the mobility minima and the rich commensurate behavior of the independently variable $\Delta_{BNS}$ BNS simulations were closely analogous to those of the NP-FET simulations, where the $\Delta_{FET}$ controlled the filling $FF$.

In closing, it is important to understand the commensuration effects in bilayer NP solids as in NP-FETs electron transport is confined to the first few NP layers adjacent to the substrate. In such confined spaces the Coulomb-blockade-induced commensuration effects tend to introduce profound blockades against electron transport. Such transport blockades can greatly hinder the usefulness and adoption of NP-FETs for opto-electronic applications. Our work intended to serve as a guide how to control, avoid and overcome transport blockades induced by the interplay of commensuration and Coulomb effects in NP-FETs and in bilayer NP solids.

\underline{Acknowledgements}: The authors acknowledge very helpful discussions with Matt Law and Victor Klimov. This work was supported by the UC Office of the President under the UC Laboratory Fees Research Program Collaborative Research and Training Award LFR-17-477148.

\appendix*
\section{\label{appendix}Computer simulation methods}

Here we describe some of the details of the {\bf Hi}erarchical {\bf N}anoparticle {\bf T}ransport {\bf S}imulator, or HiNTS. The {\it ab initio} levels (1)-(3) of the main text have been described in earlier publications in some detail \cite{carbone_monte_2013, qu2017metal}.

Here we concentrate on layer (4), the Kinetic Monte Carlo (KMC) modeling. We introduced the KMC method to calculate mobilities of size- and lattice-disordered NP arrays. The semi-classical KMC consists of tabulating possible events and then selecting and executing events using a MC-like procedure. In particular, we decided to choose the BKL algorithm\cite{Bortz197510}. In the BKL method, each time step requires drawing two uniformly distributed random numbers between 0 and 1: $r_1$ and $r_2$. 

The simulation is initialized and then the time-evolution starts by determining the rates $\Gamma$ of possible events and then in each step we find the event $j$ for which the below equation is satisfied.

\begin{equation}\nonumber
\sum_{i=1}^{j-1}\Gamma_i<r_1\Gamma_\text{sum}<\sum_{i=j+1}^N\Gamma_i
\end{equation}

\begin{equation}\nonumber
\Gamma_\text{sum}=\sum_{i=1}^{N} \Gamma_i
\end{equation}

Then event $j$ is executed and the time is advanced by drawing a second uniform random number:

\begin{equation}\nonumber
\Delta t=\frac{-\ln(r_2)}{\Gamma_{sum}}
\end{equation}

Finally, all of the events that may have changed are recalculated. Simulation is stopped when the measured physical observable reached a steady state value within a user-defined threshold.


We start the simulation by randomly placing charges on NPs with predefined density, and then we switch on the KMC algorithm. The mobility is measured as:

\begin{equation}\nonumber
\mu_e=\frac{\text{harvested charges}\times L_z}{\text{t}\times\text{total number of carriers}\times F_\text{ext}}
\end{equation}

where $L_z$ is the length of the simulation box in the conducting direction. As stated above, we stop the simulation once the mobility reaches steady state: typically millions of time steps are needed to reach convergence. We used periodic boundary conditions in all three Cartesian directions and the number of harvested charges were measured by counting the net number of electrons crossing the $z=L_z$ plane.

For the regular hopping, we use Miller-Abrahams (MA) thermally assisted nearest-neighbor hopping or tunnelling. Other approaches include the Marcus theory of electron transfer\cite{marcus_theory_1956}, which also takes into account nuclear relaxation effects after the hopping, and the model developed by Nelson and Chandler that closely resembles Marcus theory.\cite{chandler_electron_2007}

The validity and differences of these approaches have been analyzed in detail.\cite{stephan_monte_2000, vukmirovic_carrier_2010}

\begin{equation}\label{eq:ma}
\Gamma_{i\rightarrow j}= 
\begin{cases}
\nu\beta_{ij}\exp\left(\frac{-\Delta E_{ij}}{k_bT}\right)& \text{if $E_{i}>E_{j}$},\\
\nu\beta_{ij}& \text{if $E_{i}<E_{j}$}\\

\end{cases}
\end{equation}

The attempt frequency $\nu$ is assumed to be size and ligand independent and they set the time scale of the simulations. We chose $\nu$ in order to qualitatively match the order of magnitude mobitilies measured by the Matt Law group.\cite{liu2013pbse} $\Delta E_{ij}$ is the energy difference between electron states of the $ith$ and $jth$ NPs. $\beta$ is tunnelling amplitude and we evaluate it in the WKB approximation:

\begin{equation}\label{eq:wkb}
\beta_{ij}(E)=\exp\left(-2\Delta x\sqrt{\frac{2m^*(E_{\text{vac}}-E^{tunnelling})}{\hbar^2}}\right)
\end{equation}

Here $\Delta x$ is the NP-NP surface-to-surface distance, which in practice is chosen to be twice the ligand length. $m^*$ is the effective mass of the tunnelling medium, which also depends on the effective mass of the barrier. Here we approximated $m^*$ with the effective masses of electrons and holes in bulk PbSe.\cite{pbse-dielectric-constant} An alternative approach is to use the Bardeen formula of tunnelling.\cite{lepage2013modelisation} Here we refer everything to the vacuum level $E_{\text{vac}}$ which is thus set to zero in all simulations. If the NP solid was embedded in a matrix this would represent the conduction band minimum of the embedding matrix. $E^{tunnelling}$ is the tunnelling energy. It is not immediately clear what energy should be used for $E^{tunnelling}$. In the spirit of the thermally assisted hopping approach of Chandler and Nelson, $E^{sp}$ was defined as an average of the energies of initial and final states of the hopping: $E^{tunnelling}=(E^{sp}_a+E^{sp}_b)/2$, where $E^{sp}$ is the single particle energy.

The energy difference $\Delta E_{ab}$ in Eq.\ref{eq:ma} is the barrier for hopping, which can be written as:

\begin{equation}\label{eq:energydiff}
\Delta E_{ab}=\Delta E_{ab}^{sp}+\Delta E_{ab}^{F}+\Delta E_{ab}^{C},
\end{equation}

where the first term on the RHS is the difference in single particle energies of the initial and final states of the hopping:

\begin{equation}\label{eq:sp}
\Delta E_{ab}^{sp}=E_b^{sp}-E_a^{sp}.
\end{equation}

We used the energies from $k \cdot p$ perturbation theory as obtained by Kang and Wise\cite{kang_electronic_1997}. We then applied a rigid shift to align the infinite diameter limit of the conduction band edge to the work function of bulk PbSe. \cite{jasieniak_size-dependent_2011}

The second term on RHS of Eq.~\ref{eq:energydiff} is the contribution from the external voltage $V$:

\begin{equation}
\Delta E_{ab}^{F}=q\frac{V}{L_z}(z_b-z_a),
\end{equation}
and
\begin{equation}
\Delta E_{ab}^{F}=\Delta
\end{equation}

where $L_z$ is the length of the NP solid in the conducting direction and $z_b$ and $z_a$ are the $z$ position of the center of the NPs, and $\Delta$ is the energy difference associated with the transverse field. Care is exercised to make sure simulations are always in linear I-V regime. In particular, we set the external voltage so that $|E^{F}_{ab}|=0.1kT(@30K)$. As mentioned in the main text, the disorder of the NP energies did not exactly average to zero in our samples, and thus generated an internal bias field. We eliminated this bias by always taking the pairwise average of the currents with a forward and a backward applied voltage.

Finally, the last term is due to the on-site Coulomb interaction:

\begin{equation}
\Delta E_{ab}^{C}=\Sigma_b^0+(n_b)\Sigma_b-(\Sigma_a^0+(n_a-1)\Sigma_a)
\end{equation}

where we introduced self energy, or the on-site charging energy: $\Sigma^0$ is the energy that needs to be paid upon the load of the first charge onto the neutral NP, while $\Sigma$ is the energy it takes to load each additional charge. Both of them can be written in the form of $\Sigma(d_\text{particle})=q^2/2C(d_\text{particle})$, where is $C$ is the self-capacitance of the NP. Some groups also include here the mutual capacitance of the array further decreasing the self-energy.\cite{liu_dependence_2010} The capacitance can be taken to be proportional to the diameter $d$. This is the approach we followed in our previous work and $X_C$ was chosen according to the work of Zunger\cite{an_electron_2007}. In this work, instead, we use Delerue's model\cite{PhysRevLett.84.2457}, which provides a semi-analytic form for $\Sigma$ and for $\Sigma_0$:

\begin{equation}\label{eq:sigma0}
\Sigma^0 = \frac{q^2}{8\pi \epsilon_0 R}\left(\frac{1}{\epsilon_{\text{solid}}}-\frac{1}{\epsilon_{\text{NP}}}\right)+\frac{0.47q^2}{4\pi \epsilon_0\epsilon_{\text{NP}} R}\frac{\epsilon_{\text{NP}}-\epsilon_{\text{solid}}}{\epsilon_{\text{NP}}+\epsilon_{\text{solid}}}
\end{equation}

\begin{equation}\label{eq:sigma}
\Sigma= \frac{q^2}{4\pi \epsilon_0 R}\left(\frac{1}{\epsilon_{\text{solid}}}+\frac{0.79}{\epsilon_{\text{NP}}}\right)
\end{equation}

We used the Maxwell--Garnett (MG) effective medium approximation\cite{753004} to compute the dielectric constant of the entire NP solid. According to MG, the dielectric constant of the solid can be approximated as

\begin{equation}\label{eq:mg}
\epsilon_\text{solid}=\epsilon_\text{ligand}\frac{\epsilon_\text{NP}(1+\kappa f)-\epsilon_\text{ligand}(\kappa f-\kappa)}{\epsilon_\text{ligand}(\kappa+f)+\epsilon_\text{NP}(1-f)}
\end{equation}

where $\kappa$ is 2 for spherical NPs and $f$ is the filling factor.

Determining the dielectric constant of NPs is a field on its own\cite{PhysRevLett.73.1039,PhysRevLett.94.236804}. The high frequency dielectric constant of bulk PbSe is 22.9 at room temperature, while the low frequency dielectric constant is 210 at room temperature.\cite{pbse-dielectric-constant} It is not immediately clear whether the dielectric constant entering Eqs.~\ref{eq:sigma0},\ref{eq:sigma},\ref{eq:mg} should contain ionic relaxation effects or not. Furthermore, the dielectric constant of a single NP is in principle one by definition. One can usually define an effective dielectric constant if the NP is big enough\cite{PhysRevLett.73.1039,PhysRevLett.94.236804} but it turns out that such models, e.g. Penn Model\cite{PhysRev.128.2093}, may not necessarily work for any kind of system.\cite{pan2014refractive} In order to avoid making an uncontrolled approximation we decided to use the high frequency bulk dielectric constant of PbSe in the entire NP diameter range.\cite{an_electron_2007}

Having defined the energetics of the NPs we can now discuss the transition, or hopping rates. The probability of an electron transferring from the initial NP $a$ to the NP $b$ is:
\begin{equation}
\Gamma_{ab}=\sum_{ij}\Gamma_{ij}g_if_i(n_a)g_j(1-f_j(n_b))
\end{equation}

where $i$/$j$ denote kinetic energy levels, $g_i$/$g_j$ are their degeneracy and $f_i$/$f_j$ is the Fermi occupation function, $n$ is the number of electrons on the respective nanoparticle. 

Since we are assuming that the NP solid is weakly charged and the average charge per nanoparticle is less than the band degeneracy, the Fermi occupation functions can be replaced by their zero temperature limit and the sum over bands is limited to the first states:

\begin{equation}
\Gamma_{ab}=\sum_{\substack{i\in {occ} \\ j\in \text{unocc}}}\Gamma_{ij}g_ig_j
\end{equation}

Following earlier works and assuming that our NP solid is not operating in the extremely confined size regime, the degeneracy $g$ of the band edge states comes from the valley-degeneracy of bulk PbSe which is eight, including spin. Later work of Delerue showed that there is minor split of these states due to intervalley coupling.\cite{allan_confinement_2004}


In order to investigate more realistic lattice disordered NP solids we set up random closed packed models by using an event-driven Molecular Dynamics code.\cite{Donev2005737,Donev2005765} NP diameters ($d$) were drawn from a Gaussian distribution with an average diameter $\mu$ and standard deviation of $\sigma$:

\begin{equation}
f(d, \mu, \sigma) = \frac{1}{\sigma\sqrt{2\pi}} e^{ -\frac{(d-\mu)^2}{2\sigma^2} }
\end{equation}

In order to sufficiently capture disorder effects we averaged over one hundred different random NP lattices. Error bars in our calculations represent the standard deviation of the mean. 

\input{main.bbl}

\end{document}

%% file: main.bbl
%